\documentclass[useAMS,usenatbib,a4paper]{mn2e}
\input{psfig}
\usepackage{amsfonts,amssymb,amsmath}
\usepackage{aas_macros}
\usepackage{times,varioref,color,multirow,textcomp}
\usepackage[
    dvips,
    a4paper=true,
    plainpages=true,
    pdfpagelabels,
    bookmarks=true,
    bookmarksopen=false,
    bookmarksopenlevel=2
    bookmarksnumbered=true,
    bookmarkstype=toc,
    colorlinks=true,
    citecolor=red,
    linkcolor=blue,
    menucolor=green,
    urlcolor=magenta,
]{hyperref}
  
\begin{document}


\title[Galaxy clustering in semi-analytical models]
      {Resolving the problem of galaxy clustering on small scales: any new physics needed?}
\author[X. Kang etal.]
       {X. Kang$^{1}$\thanks{E-mail:kangxi@pmo.ac.cn}\\
        $^1$Purple Mountain Observatory, the Partner Group of MPI f\"ur Astronomie, 2 West Beijing Road, Nanjing 210008, China}


\date{}

\pagerange{\pageref{firstpage}--\pageref{lastpage}}
\pubyear{2013}

\maketitle

\label{firstpage}


\begin{abstract}

Galaxy  clustering sets  strong constraints  on the  physics governing
galaxy formation  and evolution. However, most current  models fail to
reproduce  the  clustering  of   low-mass  galaxies  on  small  scales
($r<1Mpc/h$).  In this paper we study the galaxy clusterings predicted
from  a few  semi-analytical models.   We firstly  compare  two Munich
versions,  Guo et al.  (2011, Guo11)  and De  Lucia \&  Blazoit (2007,
DLB07).  The  Guo11 model  well  reproduces  the  galaxy stellar  mass
function,  but over-predicts  the clustering  of low-mass  galaxies on
small scales.  The DLB07 model provides a better fit to the clustering
on small  scales, but over-predicts  the stellar mass  function. These
seem to be  puzzling.  The clustering on small  scales is dominated by
galaxies  in the same  dark matter  halo, and  there is  slightly more
fraction of satellite galaxies residing in massive haloes in the Guo11
model,  which   is  the   dominant  contribution  to   the  clustering
discrepancy  between  the  two  models.  However,  both  models  still
over-predict  the  clustering   at  $0.1Mpc/h<r<10Mpc/h$  for  low  mass
galaxies.   This is  because both  models over-predict  the  number of
satellites by  30\% in massive halos  than the data. We  show that the
Guo11  model could  be  slightly modified  to  simultaneously fit  the
stellar  mass function  and clusterings,  but that  can not  be easily
achieved in the DLB07 model.  The better agreement of DLB07 model with
the  data actually  comes as  a coincidence  as it  predicts  too many
low-mass central galaxies which are less clustered and thus bring down
the  total clustering.   Finally,  we show  the  predictions from  the
semi-analytical of  Kang et al. (2012).   We find that  this model can
simultaneously fit the stellar  mass function and galaxy clustering if
the supernova feedback in satellite galaxies is stronger.  We conclude
that  semi-analytical models are  now able  to solve  the small-scales
clustering  problem, without  invoking  of any  other  new physics  or
changing the dark  matter properties, such as the  recent favored warm
dark matter.

\end{abstract}

\begin{keywords}
methods: analytical --
galaxies: mass function -- 
galaxies: formation --
cosmology: theory -- dark matter -- large-scales structure of Universe
\end{keywords}


\section{Introduction}
\label{sec:intro}

In the cold dark matter  universe, structure formation is dominated by
dark matter haloes,  and their formation and distribution  can be well
studied using high-resolution N-body  simulations (e.g., Navarro et al.
1997; Springel et al.  2005a; Li et al.  2012). However, the formation
of galaxies  involves baryonic process  which are much  more uncertain
and complicated.   To understand  and constrain how  galaxy population
form and distribute in a statistical  point of view, one often needs a
few important  observables: galaxy luminosity/stellar  mass functions,
clusterings, and  color distributions.  Luckily,  local galaxy surveys,
such as the  Sloan Digital Sky Survey (SDSS, York  et al.  2001), have
accurately measured these observables in the last decade. They are now
widely used  as inputs to  constrain the models for  galaxy formation:
such as the Halo occupation  distribution (HOD, e.g., Peacock \& Smith
2000; Seljak  2000; Ma \&  Fry 2000; Kang  et al.  2002;  Cooray 2002;
Zheng et  al.  2005), the conditional luminosity  function (CLF, e.g.,
Yang et al.  2003; van den Bosch et al.  2007), the abundance matching
method (e.g., Vale  \& Ostriker 2004; Conroy \&  Wechsler 2009; Moster
et al.  2010), and the semi-analytical models (SAMs, e.g., Kang et al.
2005;  Croton et  al.  2006;  Bower et  al.  2006;  Somerville  et al.
2008; Guo et al. 2011).

Among these models  or techniques, SAMs are especially  useful as they
include baryonic physics regulating star formation process. Unlike the
HOD models which often take both the stellar mass functions (hereafter
SMFs)  and galaxy  clustering as  inputs,  the parameters  of SAM  are
usually tuned  to fit  the local SMF  or luminosity  functions.  Other
observables,  such as  galaxy clustering  and color  distribution, are
seen as model predictions.  Recent years have witnessed great progress
in achieving better  agreement with the data from  the SAMs.  However,
though they  can now well reproduce many  observables separately, most 
of them  are unable  to reproduce the  SMFs, galaxy clustering  and color
distribution  simultaneously  (However  see  recent progress  made  by
Henriques et al. 2013).  For  example, the recent models (Bower et al.
2006; Guo et al.  2011, here  after Guo11; Kang et al. 2012, hereafter
K12) can well  reproduce the measured local SMFs perfectly  (Cole et al.  2001;
Bell  et al.   2003;  Li \&  White  2008), but  they over-predict  the
clusterings on small scales.  The  model of De Lucia \& Blaizot (2007,
hereafter DLB07)  over-predicts the local SMF, but  is more successful
in reproducing  galaxy clustering (Wang et al.   2013). By introducing
gradual strangulation of hot halo  gas of satellite galaxy, SAMs (Kang
\&  van den  Bosch  2008; Font  et  al 2008)  can  also reproduce  the
observed color  distribution of satellites from the  SDSS (Weinmann et
al.  2006; van den Bosch et al. 2008).

It seems  to be quite puzzling  and desperate that SAMs  are unable to
reproduce these important  observables simultaneously.  In particular,
the failure to reproduce galaxy clustering on small-scales ($<1Mpc/h$)
has intrigued  a few attempts  to modify the  cosmological parameters
and dark matter properties. However, as  found by a few recent works (
K12; Guo et al. 2013),  SAMs still over-produce galaxy clustering in a
lower $\sigma_{8}$ universe  other than the WMAP1 one  (Spergel et al.
2003).  Kang et  al.  (2013) found that a  warm dark matter cosmology,
after  tuning  the  model  parameter  to  fit  the  local  SMF,  still
over-predicts clusterings of low-mass galaxies on small scales.

Recently, Wang et al.   (2013, hereafter Wang13) carefully investigate
the origin for the discrepancy  between the two versions of the Munich
models, namely the Guo11 and DLB07 ones.  Both models adopted the same
dark matter  merger trees from the Millennium  Simulation (Springel et
al.   2005a).   They also  share  very  similar  descriptions for  the
baryonic  process of star  formation.  Slight  differences are  in the
treatments  of  supernova feedback,  gas  cooling  of satellites,  
satellite  disruption, and etc.   Wang13  found  that the  scatter  around  the
stellar mass to halo mass relation can explain the discrepancy between
the two  models.  They claimed  that galaxies above the  mean relation
form earlier,  and this effect is  stronger in the  Guo11 model.  They
thus concluded  that this formation  bias accounts for  the clustering
discrepancy between the two models.

The work of Wang13 provides an  useful insight into the origin for the
clustering  discrepancy between  the  Guo11 and  DLB07 models.  Wang13
indicates  that the  halo formation  bias  (or assembly  bias) is  the
reason  for the  difference of  clustering  on small  scales.  It  was
previously recognized that  the halo assembly bias has  an effect only
on large scales (e.g., Gao et al.  2005; Wechsler et al. 2006; Jing et
al. 2007). However,  it is only recently found  that the assembly bias
also accounts for the mass  distribution on small scales, such as halo
concentration, subhalo fraction. (e.g., Gao et al. 2007). Thus in this
paper we investigate  this problem in more detail.   For that purpose,
we use the public available data from the German Astrophysical Virtual
Observatory (Lemson \& the Virgo Consortium 2006).

We  compare  the  properties  of  galaxies  from  the  Munich  models,
including their host halo mass distribution, galaxy density profile in
the  host halo,  and the  conditional stellar  mass  functions (CSMFs:
stellar mass function  in host haloes with given  mass).  We find that
the main  reason for the  discrepancy of clustering between  the Guo11
and DLB07  models is not from the  halo assembly bias, but  due to the
fraction  of satellite galaxies  in massive  haloes. Also  both models
over-predict  the   CSMFs  in  massive   haloes  by  around   30\%  at
$M_{\ast}=10^{10}M_{\odot}$.  We discuss a  few methods to rescale the
models  to best  fit the  global SMF  and CSMFs,  and show  that after
fixing  the  match  to  the   SMFs,  the  galaxy  clustering  is  also
reproduced.

In addition to  the investigation of the Munich  models, we also study
the predictions from  the K12 model.  This model  also well reproduces
the local  SMF by  introducing a  lower gas cooling  rate in  low mass
haloes compared  to its previous version  (Kang et al.   2005; Kang \&
van den  Bosch 2008).  However  it also predicted higher  clustering on
small  scales. We  find that  it is  mainly due  to  the unreasonable
description  of supernova  feedback  in  satellite galaxies.  We  slightly
modify the  efficiency of supernova  feedback, and find that  the new
model  can now  well reproduce  the global  SMF and  CSMFs.   Also the
galaxy clusterings are now well reproduced on all scales.

The  paper is  organized as  follows. In  Sec.~\ref{sec:model-mpa}, we
briefly   introduce the  model implementation  for the  DLB07  and Guo11
models, and show their model predictions. In Sec.~\ref{sec:model-K12},
we show the modification to the K12 model with compare the predictions
to  the  data.  We  give  our  conclusion  and simple  discussions  in
Sec.~\ref{sec:cons}.

\section{Munich semi-analytical models}
\label{sec:model-mpa}

\subsection{Description of the Model}

The  Munich model  is mainly  based  on the  papers by  White \&  Rees
(1978), White \&  Frenk (1991), and Kauffmann et  al.  (1999).  It was
improved with the  inclusion of subhaloes by Springel  et al.  (2001),
and a  model for  suppression of  cooling flow by  'Radio AGN'  in the
model of  Croton et al.  (2006).   The DLB07 model is  very similar to
that  of  Croton,  with  modifications  to the  stellar  initial  mass
function and more realistic dust  model, and use of slightly different
merger trees.  However, these  previous models over-predict the SMF at
$z=0$.  Further  improvement is implemented  in the recent  version of
Guo11.  In  this new  model, they include  a more  efficient supernova
feedback  and different  treatments  of satellite  evolution, such  as
allowance of gas cooling in  satellites, gradual stripping of hot halo
gas  and satellite disruption.   The Guo11  model well  reproduces the
local stellar mass function from the SDSS data (Li \& White 2008), but
over-predicts  the  clustering  on  small  scales  (Wang13).   In  the
following,  we list  the main  physical implementations  in  the Guo11
model which are not included in  the DLB07 model. For more details, we
refer the readers to the papers of DLB07 and Guo11.

\begin{figure*}
\centerline{\psfig{figure=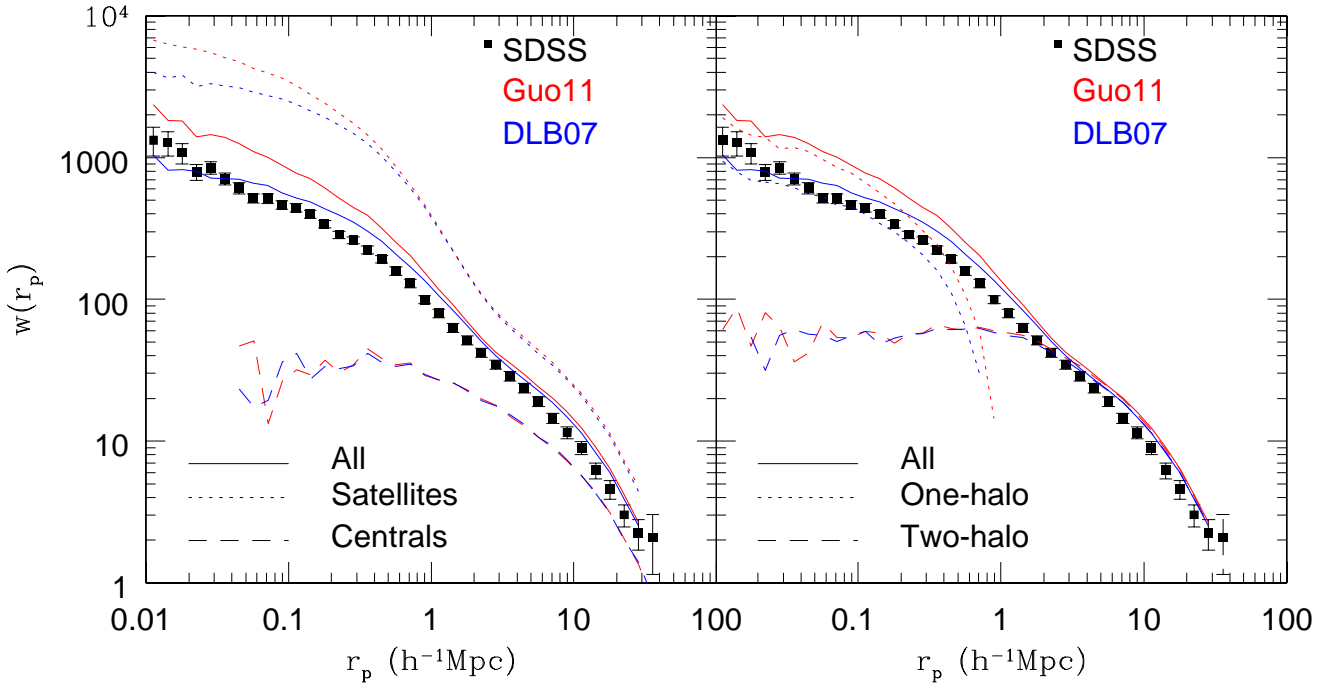,width=0.98\textwidth}}
\caption{Projected two-point correlation function of galaxies. The data points are from Li et al. (2006). Red (blue) solid lines are for Guo11 (DLB07) models. Left panel: The  dotted and dashed lines are the auto-correlation of  satellites and central galaxies. Right panel: the contribution from the one-halo and two-halo terms.}
\label{fig:2pcf-mpa}
\end{figure*}

The  Guo11  model  introduces  a  few modifications  and  new  physics
prescriptions compared  to the DLB07  model.  The first is  that Guo11
implemented a  different description for gas reheat by supernova  feedback, 
which is dependent on the gravitational potential and mainly  affects the star  formation efficiency  in low  mass haloes.   The second  is that
satellite galaxies in  Guo11 model could have more  gas cooling as its
host  halo gas  is now  gradually stripped,  unlike  the instantaneous
stripping adopted  in the DLB07  model.  Thirdly, satellite  galaxy in
the Guo11 model will be disrupted if the baryonic density of satellite
is  less  than  the dark  matter  density  of  the  host halo  at  the
pericenter.  The  tidal  disruption  is  not  included  in  the  DLB07
model. Another modification in the  Guo11 model is that the assignment
of  position for  orphan galaxy  (defined as  that  without associated
subhalo,  Gao et  al.  2004) is  different.  In the  DLB07 model,  the
position of an orphan galaxy is tagged by tracer particle (the particle
with the  most bound  energy at  the time when  its subhalo  is lastly
resolved).  Guo11  assume that the  distance of a satellite  galaxy to
the host halo  center, $R_{new}$, is not given  by the tracer particle
as $R_{tracer}$, but scaled  as\footnote{There was a typo in the original formula of Guo11 (private communication).},
\begin{equation}
\Delta R_{new} = (1-\Delta t/t_{friction})^{0.5} \Delta R_{tracer} 
\end{equation}
where $\Delta  t$ is the time  since the merger clock  of satellite is
reset  when its  associated  subhalo  is not  resolved  any more,  and
$t_{friction}$ is the dynamical friction time which indicates how long
it will take for the satellite to merger with its central galaxy.

As shown by Wang13, the effect  of the first modification in the Guo11
model  is that  the star  formation  in low-mass  halo is  suppressed,
leading to  fewer low-mass galaxies, thus fitting  better with the
global SMF.   The introduction of satellite  disruption also decreases
the number of satellites, and it agrees better with the CSMFs (we show
these  in  Fig.~\ref{fig:CSMFs}).   The  new assignment  of  satellites' 
 positions leads to  a slightly steeper density profile  of galaxies (we
will show its effect in Fig.~\ref{fig:2pcf-profile}).

The      Munich      galaxy      catalogue     is      now      publicly 
available\footnote{http://gavo.mpa-garching.mpg.de/MyMillennium}.  For
each  galaxy,  one  can  obtain  its  stellar  mass,  multi-wavelength
magnitudes,  position and  velocity.  The  database also  includes the
information of the host halo  for each galaxy, including halo mass and
formation history.   In this  paper, we use  the galaxy  catalogues of
DLB07  and  Guo11 models  based  on  the  Millennium Simulation  (Springel  et
al. 2005a).  This simulation uses  the first-year WMAP  cosmology with
$\Omega_{m}=0.25,\Omega_{\lambda}=0.75,  \sigma_{8}=0.9$  (Spergel  et
a. 2003). It includes $2160^{3}$ particles in a cub box with each side
of 500 Mpc/h. Each particle has mass of $1.18\times 10^{9}M_{\odot}$, and 
Guo11 have  shown that galaxies with stellar mass above
$M_{\ast}=10^{9.5}M_{\odot}$ are well resolved in this simulation.

\subsection{Results of the Munich model}

Wang13 have  compared the predicted galaxy clusterings  from the DLB07
and Guo11  models. They found  that for massive galaxies,  both models
predictions   agree   with    the   data.    For   low-mass   galaxies
($M_{\ast}<10^{10.27}M_{\odot}$),  the predicted clusterings  from the
DLB07 model agree better with the  data on small scales, and the Guo11
model is  higher than the data by  a factor of 2  at $r<0.1Mpc/h$.  In
this section, we  explore the origin of the  discrepancy between their
predicted clusterings  on small scales.  As galaxy clustering  is mass
dependent,  we therefore select  galaxies within  a narrow  mass range
($10^{9.77}M_{\odot}<M_{\ast}<10^{10.27}$).  In the next section we will
focus on the model discrepancy with the real data.

To  begin with our  analysis, we  show in  fig.~\ref{fig:2pcf-mpa} the
projected two-point  correlation functions (2PCFs) from  the DLB07 and
Guo11      models       for      galaxies      with our selected mass.  
The  data points are from
Li et al.  (2006) measured from  the SDSS, and the solid lines are the
model predictions.  It  is found that on large  scales (at $r>1Mpc/h$)
both models agree  with each other.  On small  scales, the DLB07 model
fits better to the data, and the Guo11 model is  higher by a factor
of 2  at $r<0.1Mpc/h$.   We note  that even for  the DLB07  model, the
clustering  is still  higher than  the data  by 30-40\%  at  scales of
$0.1Mpc/h <r < 1Mpc/h$.  On  large scales ($r>1Mpc/h$) both models are
still not  perfect and higher than  the data by around  20\% (also see
fig.20 in  Guo11).  We will  later investigate their  discrepancy with
 observation in Section.~\ref{sec:rescaled}.

Fig.~\ref{fig:2pcf-mpa} further shows the clusterings of different galaxy
samples. In the left  panel we plot the auto-correlation functions
of central and satellite galaxies.  It shows that the 2PCFs of central
galaxies in  both models  have similar amplitude.   Satellite galaxies
have stronger clusterings and they dominate the total 2PCFs. The right
panel  shows the contributions from  galaxies in  the same  halo (one-halo
term) and  in different  haloes (two-halo term).  It is found  that the
total clustering on  small scales is dominated by  the one-halo term. 
The main  contribution to  the  model discrepancy between DLB07 and Guo11 
is from  their predictions on the one-halo terms.

\begin{figure*}
\centerline{\psfig{figure=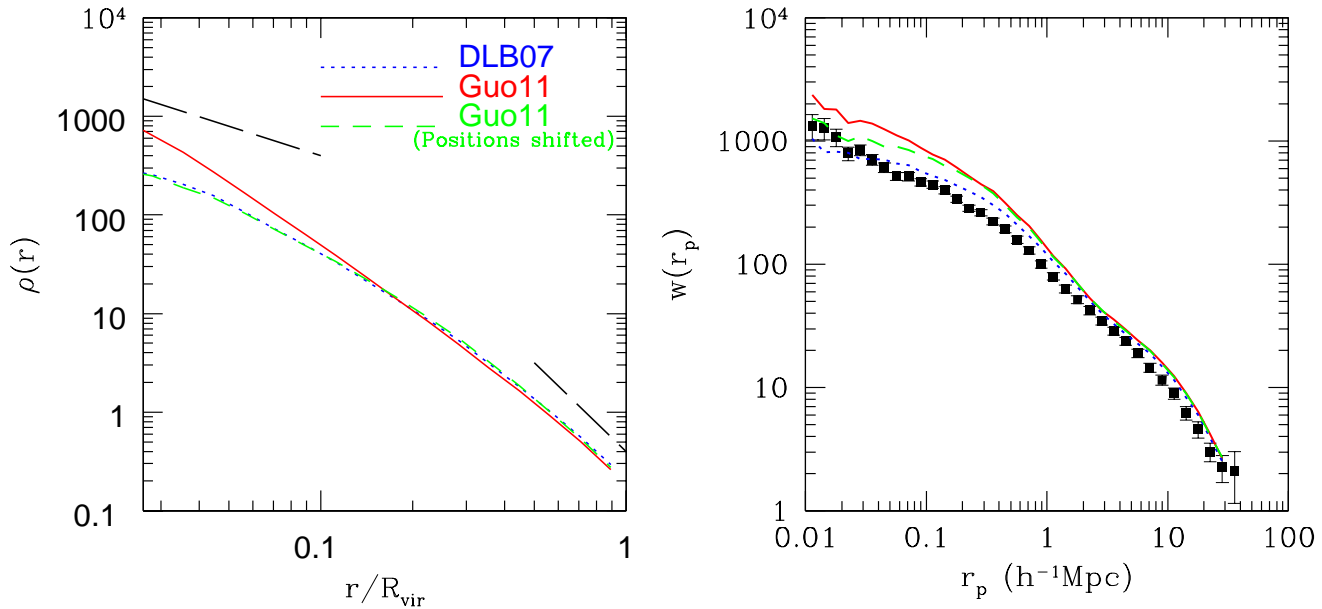,width=0.98\textwidth}}
\caption{Effects of shifting satellite positions. Left: density profiles in the DLB07 (blue dotted line) and the original Guo11 model (red solid line). Green dashed line shows the profile using the shifted positions of satellites. The long dashed line indicates the slope of a NFW profile in the inner (-1) and outer region (-3). Right: the projected auto 2PCFs of all galaxies. Shifting the positions of satellites affects the clustering only at very small scales.}
\label{fig:2pcf-profile}
\end{figure*}

Wang13 have also investigated the origin  for the discrepancy  between the
DLB07  and Guo11  models. They  found that  the  clustering difference
arises   from   the  scatter   of   the   halo   mass-  stellar   mass
relation. Galaxies  above the  median stellar mass-halo  mass relation
reside in haloes that form  earlier, while galaxies that lie below the
median  relation reside  in  haloes  form later.   Such  an effect  is
stronger in the Guo11 model. However, it  is not apparent how this effect leads
to the clustering difference at given stellar mass.  Their conclusions
seem to indicate  that the clustering difference in  the two models is
ascribed to  the formation bias  of haloes that  early formed
haloes are strongly clustered (e.g., Gao et al. 2005; Jing et al. 2007)

Our results in fig.~\ref{fig:2pcf-mpa}  do not support the argument of
Wang13, as  the halo  bias should affect  clustering more  strongly on
large  scales  (or  two-halo  terms).   Our  results  agree  with  the
prediction  from the  HOD models  that clustering  on small  scales is
dominated by the one-halo term ( e.g., Kang et al. 2002). The one-halo
term is then determined by the galaxy density profile in the host dark
matter haloes and the mass  function of host haloes (see Equation.6 in
Kang et al. 2002).  In the following we compare the predictions on the
two ingredients from the DLB07 and  Guo11 models to check which is the
dominant contribution to their clustering discrepancy.

The left panel  of fig.~\ref{fig:2pcf-profile} compares the normalized
galaxy density profiles  from the DLB07 and Guo11  models. The density
profile  is normalized  by the  total number  of galaxies  inside the
virial radius of the dark  matter haloes. The long dash lines indicate
the slopes in the inner and outer region of an NFW profile (Navarro et
al.  1997), which are -1 and -3 respectively. We find that the profile
in  the DLB07  model (blue  dotted line)  is more like  an NFW
profile, and  the Guo11  model predicts a  steeper slope in  the inner
halo (red  solid). Observations (e.g.,  Lin et al. 2004;  Budzynski et
al. 2012;) have  found that galaxy density profile  in cluster is well
described by an NFW profile  or a slightly shallower one (e.g., Adami
et  al. 1998; Sales  \& Lambas  2005).  Weinmann  et al.   (2011) also
compared  the dwarf  galaxy  profile  in clusters  to  the data,  they
claimed that  the model of  Guo11 agree better with the data,  but they
also noted that  the model prediction is still  slightly higher at 
halo center. The same conclusion  is also obtained by Guo11 themselves
by comparing galaxy profiles to the SDSS data.

Now we check whether the steeper profile in the Guo11  model is from their
assignment  on  satellite  positions.  As we stated in section.~\ref{sec:model-mpa} the position of orphan galaxy in Guo11 model is not given by the tracer particle, but shifted by a factor. Now we shift the positions in Guo11 catalogue back to those tracer particles. The  green  dashed  lines  in
fig.~\ref{fig:2pcf-profile} show the effects  of shifting the positions of
galaxies. The left  panel shows that after shifting  the positions of
orphan galaxies back to those tracer particles, as
used in DLB07,  the  density profile  now  agrees with  the DLB07  one
perfectly. This clearly demonstrates that  the steeper profile in the Guo11 model is purely  due to their rescale  of galaxy positions.  The green dashed
line in the right panel shows the predicted 2PCF. However, it is found that
the clustering is suppressed only on very small scales ($r<0.1Mpc/h$).
At  scales at  $r>0.1Mpc/h$, the  clustering from  the Guo11  model is
still higher than the DLB07 one.

\begin{figure*}
\centerline{\psfig{figure=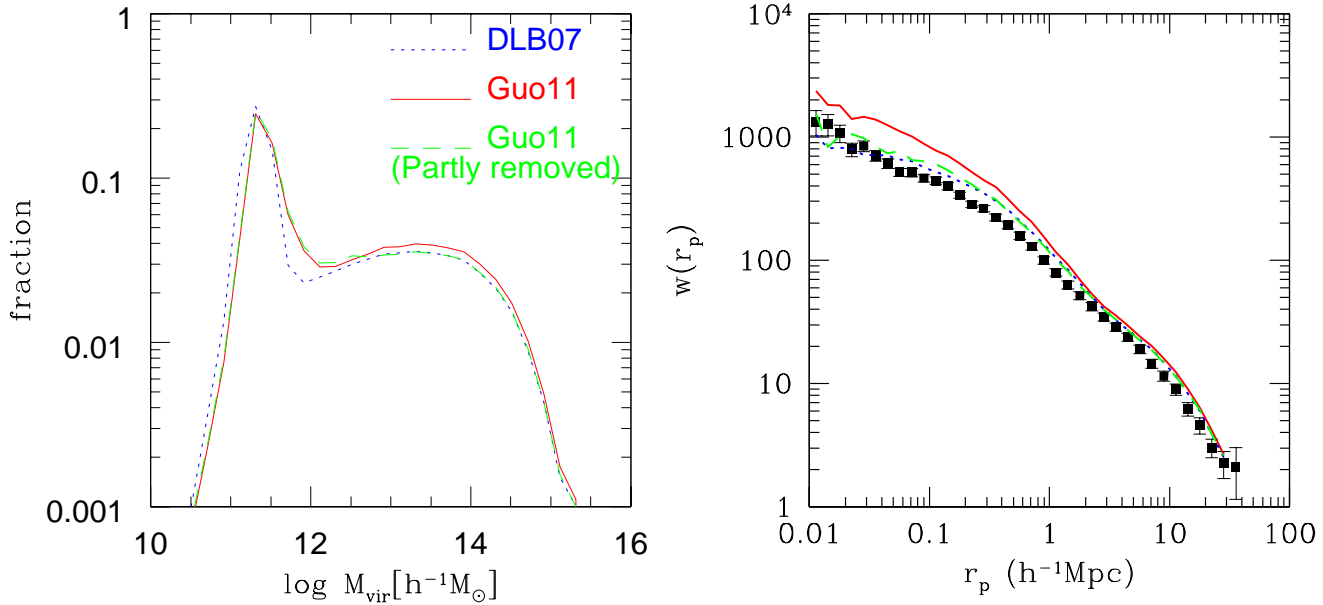,width=0.98\textwidth}}
\caption{The effect of satellites fraction on galaxy clustering. Left panel: the distributions of host  halo  mass for galaxies in the DLB07 and Guo11 models. Right panel: galaxy clusterings. The green dashed shows the effects of partly removing (by 15\%) of galaxies in massive haloes ($M_{vir}>5\times 10^{12}M_{\odot}$) in the Guo11 model. It now produces almost identical results with the DLB07 ones.}
\label{fig:fraction-mhost}
\end{figure*}

\begin{figure*}
\centerline{\psfig{figure=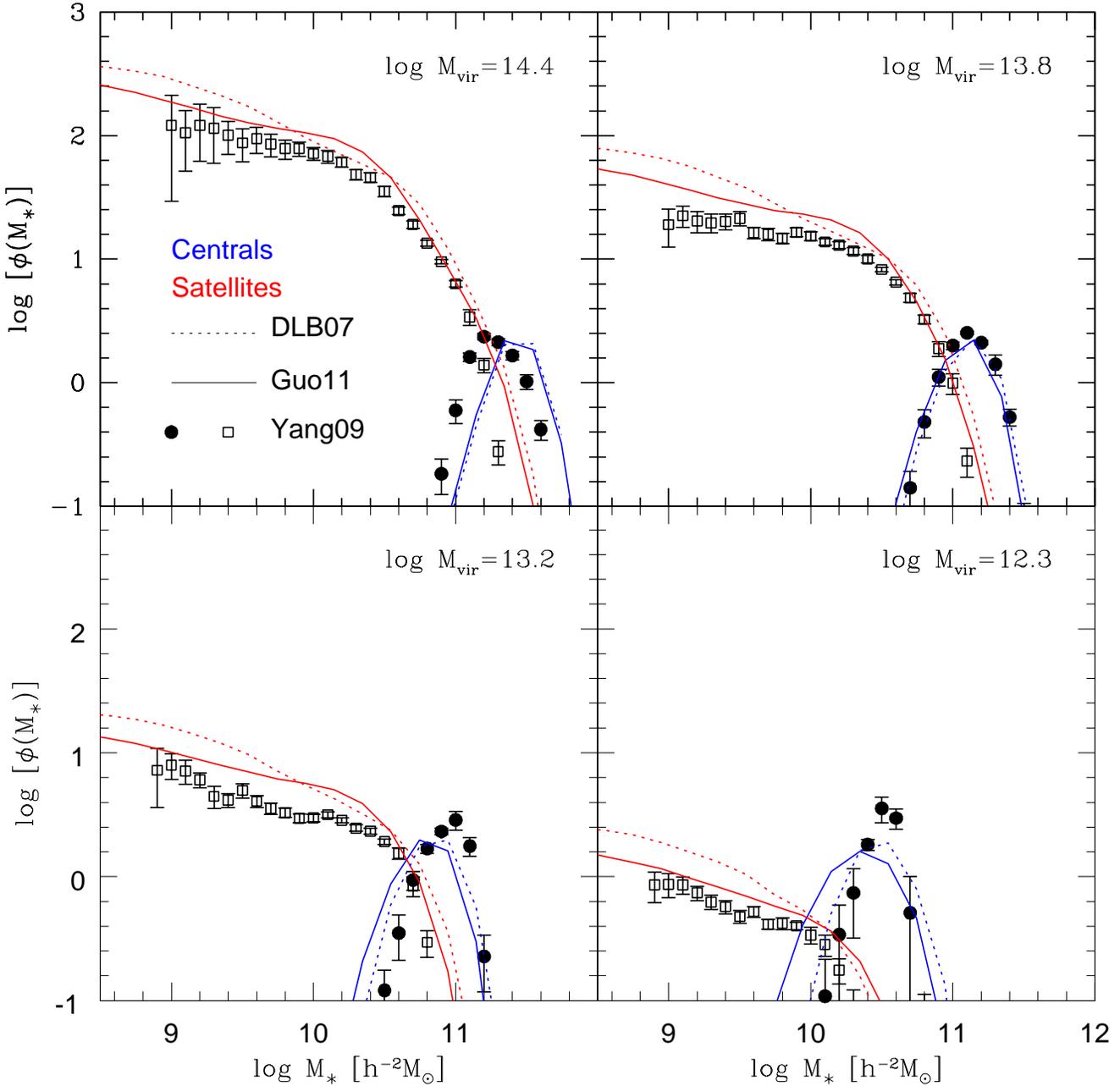,width=0.98\textwidth}}
\caption{Conditional stellar mass functions (CSMFs) in different halo mass bins. The data points are from the group catalogue of Yang et al. (2009), and red (blue) lines are for satellites (central) galaxies. Here comparisons are only shown for haloes with mass larger than $M_{vir}>2\times 10^{12}M_{\odot}$, below which the group catalogue is incomplete.}
\label{fig:CSMFs}
\end{figure*}

Fig.~\ref{fig:2pcf-profile}  indicates  that  the  difference  in  the
clusterings  predicted by  the  two  models is  not  from the  spatial
distribution of galaxies within  the dark matter haloes. Therefore the
only possible contribution is from  the mass function of the host halo
in  the  models. For  each  galaxy with  mass  in  our selected  range
($lgM_{\ast}=[9.77,10.27]$), we can obtain the virial mass of its host
halo   from   the   public   catalogue.    In  the   left   panel   of
fig.~\ref{fig:fraction-mhost} we  show the distributions  for the host
halo mass  from the two models  (DLB07: blue dotted,  Guo11: red solid
lines). At  first glance it is  found that the  distributions from the
two models  are very similar,  and there is  a sharp peak  at $M_{vir}=
10^{11.25}M_{\odot}$,  and  a  broad  distribution in  massive  haloes
($M_{vir}>10^{12}  M_{\odot}$).   It  is   easy  to  understand  that  this
distribution is from the the narrow (wide) range of the host halo mass
for central (satellite) galaxies, respectively.

Actually fig.~\ref{fig:fraction-mhost} shows that  there is  slight difference
for the  distributions of  host haloes in  the two models,  that Guo11
model has a slightly larger fraction of satellite galaxies residing in
massive haloes.  For example, we  found that the fraction  of galaxies in haloes with $M_{vir}>5\times 10^{12}M_{\odot}$ is 33.7\% in the Guo11 model,
but it is 30.3\% in the DLB07 model. Although the difference is small,
its effect on  the clustering is non-negligible. To  see its effect, we
make a simple  test that we randomly remove  some fraction of galaxies
with  $M_{vir}>5 \times 10^{12}M_{\odot}$  in the  Guo11 model so as to produce the same fraction of galaxies as that in the DLB07 model. To
achieve  that,  we  have  to  remove the  fraction  of  galaxies  with
$M_{vir}>5 \times 10^{12}M_{\odot}$ by  15\% in  the Guo11  model. The
green  dashed line in  the left  panel show  the distribution after this removal. It is found that now the fraction of  galaxies in  massive haloes  matches better that  in  the DLB07 model.

The    green     dashed    line     in    the    right     panel    of
fig.~\ref{fig:fraction-mhost}  shows the  predicted  clustering in the Guo11
model with  this removal  of satellites.  Remarkably, it  is found
that now the predicted clustering in the Guo11 model is very similar
to the DLB07  model. We note that the removed galaxies  is only 5\% of
the total selected  galaxy ($lg M_{\ast}=[9.77,10.27]M_{\odot}$). This
plot shows  that satellite galaxies in massive  haloes contribute most
to the clustering on small  scales, and a small fraction of satellites
will  produce non-negligible  effect. This  is because  the clustering
from the one-halo term is  proportional to $\rho(r)^{2}$, seen from the
equation.6 in Kang et al. (2002).

Our results above  have clearly shown that galaxy  clustering on small
scales  is dominated  by galaxies  residing  in the  same dark  matter
halo. It is found that the  steeper density profile of galaxies in the
Guo11 model contributes  to its higher clustering only on  very small scales,
and  the dominant  contribution  to the  discrepancy  between the models  
is  from the fraction of satellite  galaxies  in massive  haloes.  There  are
slightly more fraction of satellites residing in massive haloes in the
Guo11 model, leading  to a higher clustering on small  scales. It is not
clear why  there are  more satellites in  massive haloes in  the Guo11
model. One possible  reason is that the infall  halo mass of satellites
is  slightly  larger than  that  from the  DLB07  model  (see Wang13), 
and  on average only massive haloes  have accreted subhaloes
with  higher infall  mass.   Another possibility  is  that Guo11  have
introduced  tidal  disruption   for  satellites,  and  the  disruption
efficiency may be  higher in lower-mass host haloes  as they formed at
early    times    and   satellites    are    more    likely   to    be
disrupted as they have orbited in the host halo (especially in inner region) 
for longer time.  Unfortunately,  due  to   the  hardness  to  extract  halo
formation  information from  the Munich  catalogue, we  are  unable to
determine which leads to the  higher fraction of satellites in massive
haloes in the Guo11 model.

\subsection{Rescaled SAMs}
\label{sec:rescaled}
  
In the above  section, we have explored the  origin for the discrepancy
between  the predicted  galaxy  clustering from  the  DLB07 and  Guo11
models.   However,   we  find  that even after   eliminating  their  model
discrepancy, both models still predict higher clustering than the data
at large  scales at  $r>1Mpc/h$.  In this  section, we focus  on the
comparison  between  the  model and  the  data,  and  in
particular, we investigate if we can achieve better agreements with the
clustering data by rescaling their models.

Guo11 have made great progress to achieve better results on the galaxy
stellar  mass  functions  compared  to  the  DLB07  one.  However,  the
agreement with the data is still not perfect. The fig.1 of Wang13 have
shown that the global SMF is  over-predicted by about 10\% and 60\% at
$M_{\ast}=10^{10}M_{\odot}$   by   the   Guo11   and   DLB07   models,
respectively.  We need  to check  where these  over-predicted galaxies
come  from. In  fig.~\ref{fig:CSMFs} we  show the  conditional stellar
mass functions (CSMFs)  in host haloes with different  mass bins.  The
solid and dotted lines are for the Guo11 and DLB07 models, and the red
(blue) lines for satellites (centrals).   The data points are from the
group catalogues constructed by Yang et al.  (2009) from the SDSS DR4.

Overall, there are marginal agreement  between the model and the data.
Better agreement  is seen for  central galaxies on  average.  However,
the  mass of  central  galaxies in  massive  haloes is  over-estimated
(upper left panel), but under-estimated in low mass haloes, especially
for the Guo11 model (lower right  panel). The same effect is also seen
from the  K12 model  (fig.~\ref{fig:CSMFs-K12}). However, as  shown by
Wang13, the  predicted stellar mass  to halo mass relation  of central
galaxy from  the Munich model agrees  with the data  from weak lensing
and satellite kinematics  (e.g., Mandelbaum et al.  2007;  More et al.
2009) for both high and low-mass galaxies.  This is puzzling and we do
not know what  causes this discrepancy between the data  of Yang et al
and the models.   One possibility is that the  estimation of halo mass
in the data is different from  that in the simulations. This is beyond
the limits of our work, and we do not go into the details.

\begin{figure*}
\centerline{\psfig{figure=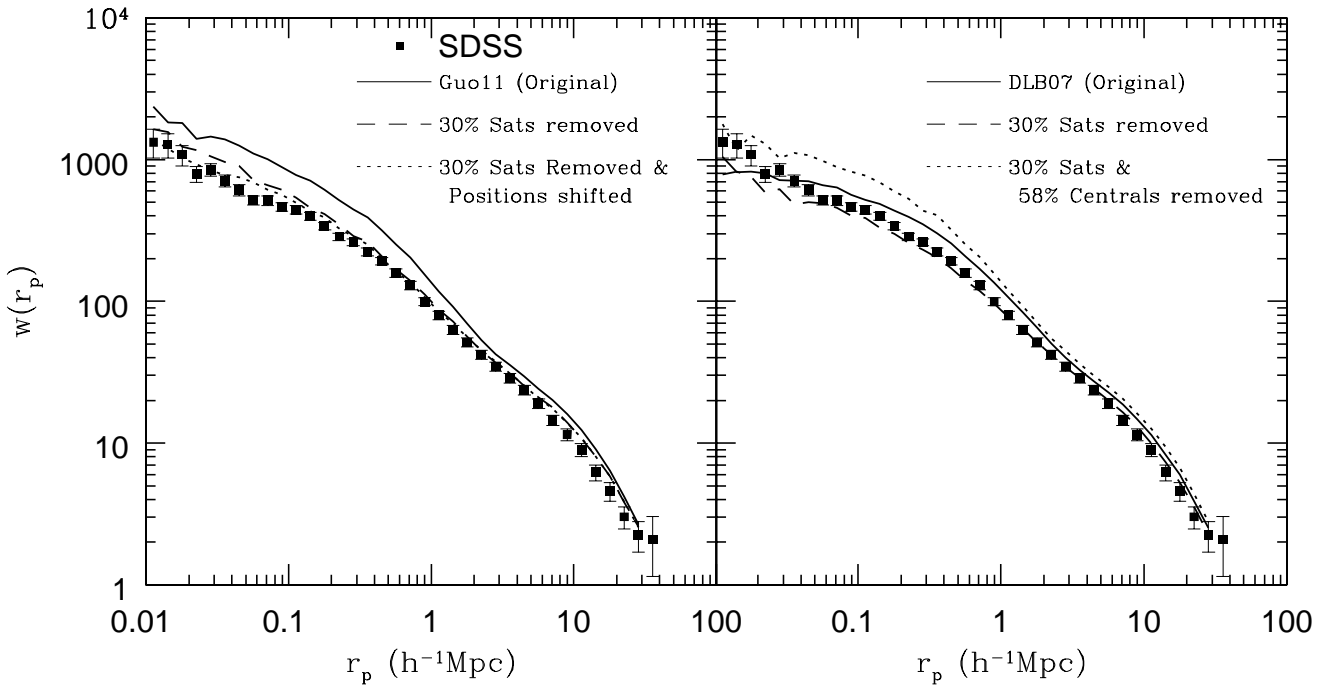,width=0.98\textwidth}}
\caption{Projected 2PCFs of all galaxies in the rescaled DLB07 and Guo11 models. Left panel is for the Guo11 model and right panel for the DLB07 model. The solid lines are results from their original models. The dotted and dashed lines show the predictions from the rescaled models, see the text for details.}
\label{fig:2pcf-scaled}
\end{figure*}

Fig.~\ref{fig:CSMFs} shows that both models over-predict the number of
low-mass satellites, and worse agreement for the DLB07 model. At our selected galaxy mass, $M_{\ast}  \sim  10^{10}M_{\odot}$, both  models  predict almost  equal
number  of satellites which  are about  30\% higher  than the  data in
haloes  with mass  larger  than $10^{12.3}M_{\odot}$.  It  is hard  to
constrain the CSMFs in lower mass haloes ($M_{vir}<10^{12}M_{\odot}$) as
the  group catalogue  in  Yang  et  al is  incomplete below  this
mass. So  for a conservative  estimate, we assume that  satellites with mass around $10^{10}M_{\odot}$ are
over-predicted  by   30\%  only  in  haloes  with   mass  larger  than
$10^{12.3}M_{\odot}$. We use this as  a constraint  in the following  analysis. Note
that the higher CSMFs of  low-mass satellites do not conflict with the
claim that Guo11 model fits the local SMF.  Wang13 have shown that the
predicted  global SMF  at $M_{\ast}=10^{10}M_{\odot}$  from  the Guo11
model  does not  match  perfectly with  the  data, but  is about  10\%
higher.     The    over-predicted     number    of    satellites    at
$M_{\ast}=10^{10}M_{\odot}$  in  halo  with  virial mass  larger  than
$10^{12.3}M_{\odot}$  is  about 9\%  of  all  galaxies  with the  same
stellar  mass.   Thus  it  could be  these  over-predicted  satellites
contributing to the over-prediction of global SMF.

Now we  investigate whether better agreement  can be achieved if we rescale both models to match the  global SMF and the CSMFs.  Wang13 have
tested two simple  models to resale the DLB07 model  to fit the global
SMF. In the first case, they  randomly removed a faction of galaxies to
reproduce the  SMF, regardless of  centrals or satellites.  They found
that random  removal of  galaxies does not  change the  original DLB07
results.  This is  easy to  understand because  simply  decreasing the
density itself does not change  the clustering. In their second model,
they removed  only satellite galaxies  and found that  the small-scale
clustering  is largely  suppressed to lower  than the  data in  the DLB07
model.

Here we use a similar method  as Wang13 to rescale the Guo11 and DLB07
models. However, we do not randomly remove galaxies, but use the CSMFs
as an additional constraint. Now  for the DLB07 model, we consider two
cases.  In the  first case we randomly remove  30\% of satellites with
$M_{\ast}=10^{10}M_{\odot}$   in   haloes   with   $M_{vir}>2   \times
10^{12}M_{\odot}$ (seen from fig.4). Note that the global SMF is still
higher  than the  data in  this  case. In  the second  case we  remove
satellites as  in the  first case,  and also  remove about  58\% of
central  galaxies so as  to fit  the global  SMF. For  the Guo11
model,  removal of  30\% satellites  (around  9\% of  all galaxies  at
$M_{\ast}=10^{10}M_{\odot}$   in   haloes   with   $M_{vir}>2   \times
10^{12}M_{\odot}$) match the global SMF quite well, and we do not
have to remove any central galaxies.

The     predicted      galaxy     clusterings    are      plotted     in
Fig.~\ref{fig:2pcf-scaled}. The left panel  shows the results of Guo11
model and the right one for the DLB07 model. It is found from the left
panel  that the  scaled Guo11  model now matches  the data  quite  well on
scales $r_{p}>1Mpc/h$, but  the small scale clustering is  slightly above the
data. However, we find that if we use the shifted positions
of satellites  (dotted line), the clustering is now perfectly  reproduced on very small scales.  The  right panel  shows that  if only  30\% of
satellites is  removed in the DLB07 model (dashed line),  the agreement with the  data is also  quite  good except  at  very small  scales.  We  note that  this
conclusion is not in conflict with the result of Wang13 as they remove
many more  satellites to  fit the global  SMF, thus the  clustering is
suppressed too much.

The dotted line in  the right panel of Fig.~\ref{fig:2pcf-scaled} shows 
the second case  of rescaling the DLB07 model in  which we remove  30\%
of satellites in massive haloes and  58\% of all centrals so as to fit
the  local SMF.  It  is seen  that  compared to  other cases,  partial 
removal  of  central galaxies  does  not  reduce  the clustering,  but
increases  it  instead.  This  result  seems  to  be  surprising.  From
Fig.~\ref{fig:2pcf-mpa} we know that  the clustering of  centrals is
much lower  than the  satellites.  This is  because for  given stellar
mass, central galaxies  often live in haloes with  mass lower than the
host of the satellites. As the  halo bias is strongly dependent on its
mass,  the  effect of  central  galaxies  is  to suppress  the  global
clustering of all galaxies.

The  results  in  fig.~\ref{fig:2pcf-scaled}  show  that  if  we  have
correctly  rescaled  the  DLB07  model  to fit  the  global  SMF,  the
predicted  galaxy clustering is  higher than  the data.   However, the
scaled Guo11 model agrees quite  well with the data. It indicates that
the  better  agreement   with  data  from  DLB07  model   comes  as  a
coincidence. It is  closer to the data than the Guo11 one because  
it wrongly predicts too many centrals.

Our test  indicates that  the Guo11 model  can be further  improved by
simply introducing a slightly stronger effect of satellite disruption
in massive haloes.  However, such  an improvement may not work for the
DLB07 model.   For the DLB07  model, the star formation  efficiency in
low-mass haloes should be suppressed, otherwise the number of centrals
is too high. In that sense,  the Guo11 model is an improved version of
the DLB07 model as it already introduces stronger feedback to suppress
star formation in low-mass haloes.

\begin{figure}
\centerline{\psfig{figure=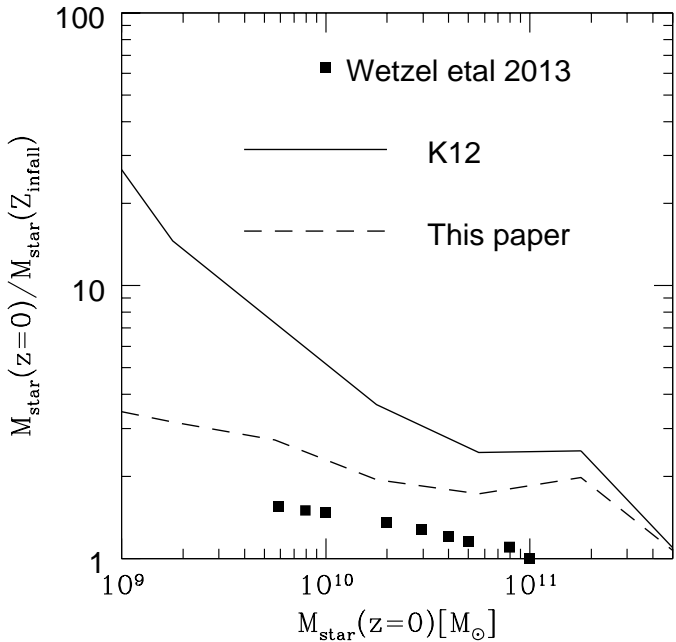,width=0.5\textwidth}}
\caption{Predicted mass growth of satellites after infall in the K12 model. The dashed line shows the result with stronger supernova feedback. Data points are from the constraints by Wetzel et al. (2013) from the SDSS DR7 and N-body simulations. }
\label{fig:satellite-growth}
\end{figure}

\section{Semi-analytical model of K12}
\label{sec:model-K12}

In  this section,  we  study  galaxy clustering  and  CSMFs using  the
slightly modified SAM of K12.  The  K12 model is based on Kang et al.
(2005;  2006; 2010)   and  Kang   \&  van  den   Bosch  (2008).    This  SAM
self-consistently models the physical processes governing stellar mass
evolution, such  as gas cooling, star formation,  supernova and active
galactic nucleus feedback.  For model details we refer  the readers to
the paper of Kang et al (2005). Compared to its previous versions, K12
introduces a  cooling factor ,$f_{c}$. For low-mass haloes, the cooling rate is then described as $\dot M_{cool}=f_{c}m_{hot}/t_{dyn}$, where $t_{dyn}$ is the halo dynamical time. As described in K12, this cooling factor takes into account the gas outflow due to reheating by supernova feedback, and it was shown to produce better match to the local SMF at the low-mass end.

The  N-body simulation used  here is  the one  presented in  K12 which
adopted  the  cosmological  parameters  from the  WMAP7  data  release
(Komatsu et al. 2011), namely: $\Omega_{\lambda}=0.73, \Omega_{m}=0.27,
\Omega_{b}=0.044$ and $\sigma_{8}=0.81$ and $h=0.7$. This simulation was run using
the  GADGET-2 code  (Springel 2005b)  in a  box of  L=200  Mpc/h using
$1024^{3}$ particles.

The K12 model has the same problem as other models that the clustering
of low-mass galaxies on small scales is over-predicted. K12 have shown
that  adopting  a  low-$\sigma_{8}$   cosmology  can  not  solve  this
problem.  They  suggested that  this  over-clustering  is  due to  the
over-prediction of low-mass satellite  galaxies in massive haloes ( or
Equally  the  masses  of  satellites  are  too  large).  The  mass  of
satellites  are too  larger  either  because they  were  too large  at
accretion  when they  were lastly  as  central galaxies,  or the  mass
growth  after  accretion  are  over  estimated. The  mass  of  central
galaxies can be constrained by the  data at $z=0$. K12 have shown that
the stellar mass - halo mass  relation of central galaxies at $z=0$ is
well  constrained by  the data  from  weak lensing.   However, such  a
direct constraint is not available at high redshift (but see Leauthaud
et al.  2012; Skibba  et al. in  preparation).  Also the  stellar mass
function at $z>0$ is also  poor constrained at low-mass end. Thus here
we do  not consider the first  possibility that the  mass of satellite
galaxies are too large at the time of accretion.

\begin{figure*}
\centerline{\psfig{figure=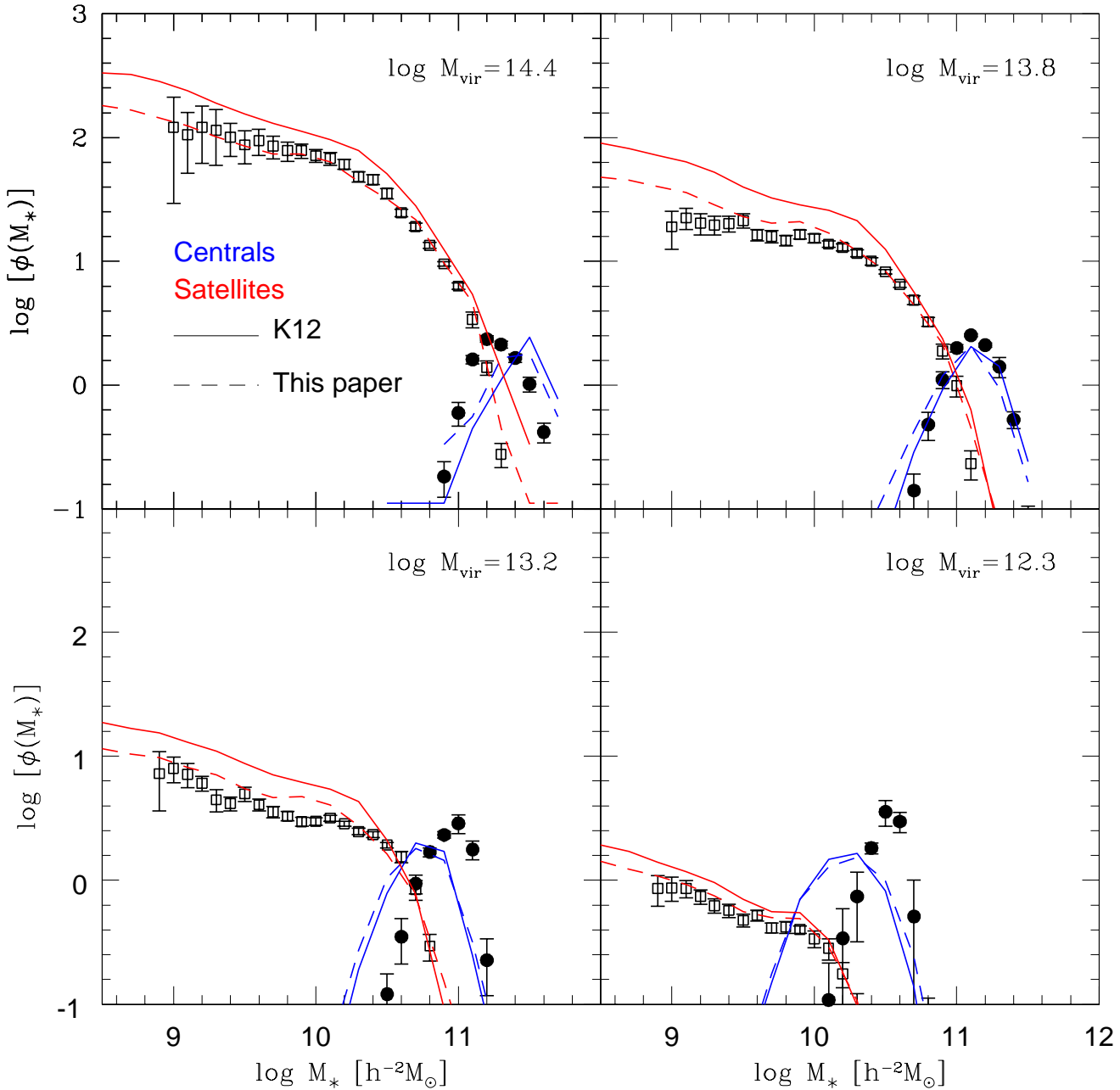,width=0.98\textwidth}}
\caption{Conditional stellar mass functions in the K12 model. The colors are the same as in Fig.~\ref{fig:CSMFs}. It is seen that stronger feedback (dashed lines) decreases the number of low-mass satellites, matching the data quite good. }
\label{fig:CSMFs-K12}
\end{figure*}

We  consider the  second possibility that  the mass  of satellites
grow too much after accretion.  In the K12 model the physics governing
satellite evolution  is gas cooling  from satellites' hot halo  and star
formation. In their model, they  considered the  impact of
supernova feedback  using energy conservation such that  the amount of
cold gas reheated by the energy from supernova is modeled as,

\begin{equation}
\Delta m_{reheated}=\frac{4}{3}\epsilon\frac{\eta_{SN}E_{SN}}{V^2_{vir}}\Delta m_{\ast},
\end{equation}
where $\epsilon$ describes the feedback efficiency, $\eta_{SN}E_{SN}$ is the energy release by supernova associated with massive stars for a unit of solar mass of newly formed stars, and $V_{vir}$ is the virial velocity of the host halo. This equation
is  under assumption  that  the cold  gas  is reheated  to the  virial
temperature  of  the  host  halo.   For satellite  galaxies,  Kang  et
al. (2005)  assumed that the cold  gas is also reheated  to the virial
temperature  of  the  host  halo.   However, this  assumption  is  not
reasonable as  supernova  feedback should be  a local  effect, and
apparently  the   satellite  galaxy   knows  nothing  about   its  host
potential. So here  we assume that cold gas is  reheated to the virial
temperature of the subhalo  it resides in, and we  use the virial velocity
of its host halo when the satellite was lastly a central galaxy before
its accretion. We  note that such a description  of supernova feedback
is also implemented in the Munich models.

\begin{figure*}
\centerline{\psfig{figure=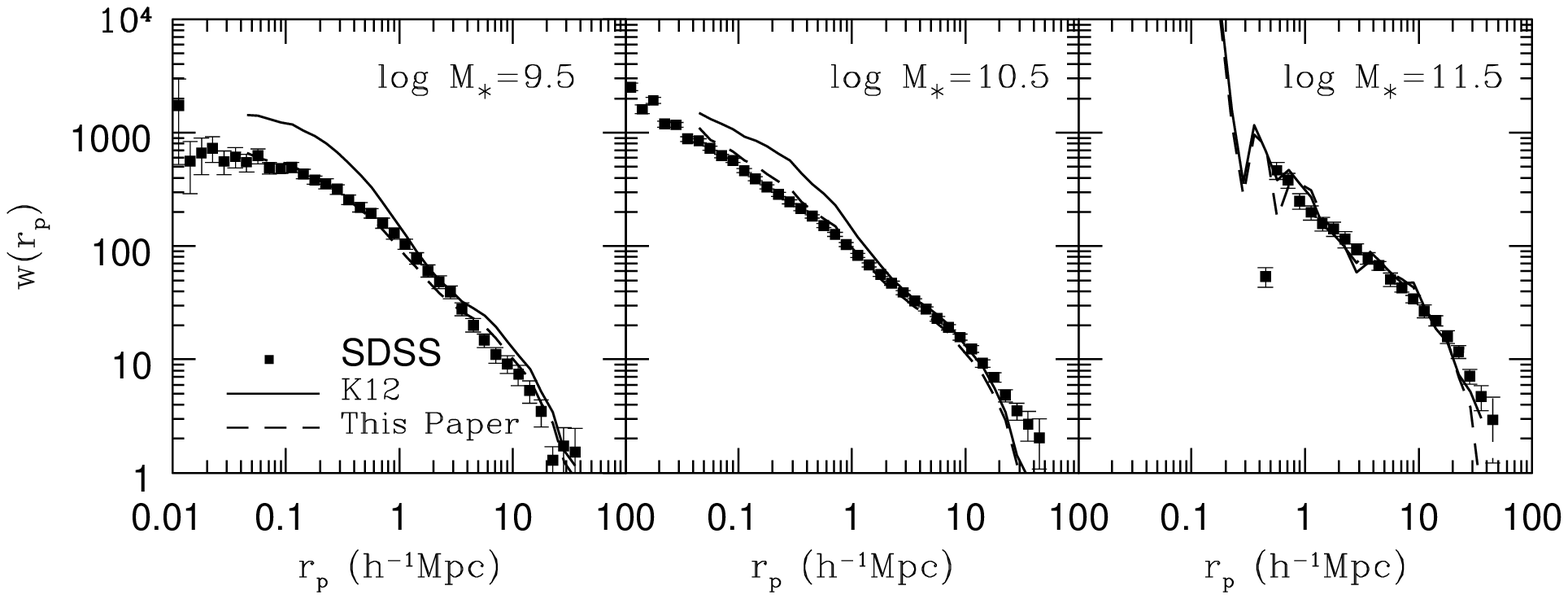,width=0.98\textwidth}}
\caption{Projected 2PCFs for low and high mass galaxies from the K12 model and its modification. Better agreement is found for the new model with stronger feedback for satellites.}
\label{fig:2pcf-K12}
\end{figure*}

Under  the above  prescription, the  supernova feedback  efficiency is
higher than the previous assumption as  the virial velocity  of satellite
galaxy is  usually lower  than its  host halo. We  keep all  the model
parameters fixed as that of K12.  As we found, the total SMF is almost
identical to  the K12 one as the  low-mass end of SMF  is dominated by
central  galaxies, and  our  modification to  the satellite  supernova
feedback has little effect on the evolution of central galaxies.

In Fig.~\ref{fig:satellite-growth}  we show the mean  mass growth rate
of  satellite  galaxies  after  accretion.   The  solid  line  is  the
prediction  from the  K12 model,  and the  dashed line  shows  the new
prediction.  It  is found  that in the  K12 model satellites  can grow
their mass  by large amount,  especially for low-mass  satellites. The
mass growth  in the new model is  largely suppressed. Observationally,
there are few direct constraints  on the mass growth of satellites. The
points are data constraints from  Wetzel et al.  (2013) using the SDSS
DR7 group  catalogue and  N-body simulations .  It shows  that stellar
mass  growth of  satellite  are typically  less  than a  factor of  2.
Similar  results on  mass evolution  of satellites  are  also recently
obtained by Watson \& Conroy  (2013). Our model result is still slightly above the data, and it is seen  that the quenching of
star formation in massive satellites is still not efficient.

Fig.~\ref{fig:CSMFs-K12} plots  the CSMFs from  the K12 model  and our
new one using the  solid and dashed lines. It is seen  that in the K12
model there are  more low-mass satellites in all  halo mass bins.  The
revised model  agrees much  better with the  data for  satellites. For
central  galaxies,  the two  model  have  similar  results and  better
agreement  with the  data is  found in  high-mass haloes.  In low-mass
haloes ($M_{vir}<10^{13.2}M_{\odot}$) the stellar mass of  central galaxies is
lower by  a factor of 2  than the data.  Our results for the  CSMFs of
central  galaxies are  very  similar  to the  Guo11  results shown  in
Fig.~\ref{fig:CSMFs}. As  we stated before, we are  not clear what
contributes to  the over-prediction  of central  galaxy mass  in high-mass
haloes and the under-prediction in  low-mass haloes. We leave this for
future work.

Now we show the predicted galaxy clusterings in Fig.~\ref{fig:2pcf-K12}
for low  and high-mass  galaxies. The solid  and dashed lines  are the
results of  K12 model and the new  one.  As it was  previously shown the
K12  model can correctly  predict the  clustering of  massive galaxies
(right  panel),  but  it  over-predicts  the  clustering  of  low-mass
galaxies on small scales (left  panels). The new model predictions are
in good agreement  with the data on all  scales.

The clustering of galaxy on small scales has attracted great interest 
recently. K12 have shown that adopting a low-$\sigma_{8}$ could
not solve this problem, and this conclusion is recently also obtained 
by Guo et al. (2013). This  is because although the number of low-mass
haloes  is  decreased  in  the  low-$\sigma_{8}$  model,  the  stellar
formation efficiency in  low-mass haloes has to be  increased so as to
fit the  local stellar mass  function. It compensates the  decrease of
subhalo in massive haloes, and still over-predicts the clustering. Besides a higher $\Omega_{m}$ adopted in a low-$\sigma_{8}$ universe will also compensated the decrease in $\sigma_{8}$.

Our results  from the new  model indicate that the  physics, governing
the evolution of satellites, such as supernova feedback or disruption,
is  crucial to  solve the  over-clustering  on small  scales. For  the
current  Guo11 model,  it can  be slightly  modified by  introducing a
stronger satellite disruption rate by about 30\% in massive haloes. As
this  percent of satellites  is only  about 10\%  of the  total galaxy
population,  it will  not violate  the agreement  on the  stellar mass
function. For the DLB07 model,  reduction of satellites alone will not do
the job  as there are  too many centrals  in low-mass haloes.  In this
model galaxies  should form in  slightly bigger haloes, so  the number
density of both  central and satellites will decrease.  In that sense,
the DLB07 model will converge with the Guo11 model.

\section{Conclusions and Discussions}
\label{sec:cons}

In  this paper, we  study the  problem of  galaxy clustering  on small
scales,  which has become  more puzzling  recently. For  this  purpose, we
utilize  the public data  from the  Munich model  with its  two recent
versions, namely Guo11  and DLB07. These two models  are very similar in 
spirit: the same  merger tree from  the Millennium  Simulation, similar
descriptions  of  gas cooling,  star formation  and  feedback from
supernova  and AGNs. However,  the Guo11  model has slightly modified 
supernova  feedback, which is  more efficient  in low-mass  haloes, and
they also  introduce satellite disruption  and use a new  algorithm to
assign the  positions of orphan  galaxies. Our results for  the Munich
models are as followings,

\begin{itemize}
 \item Although the Guo11 model  fits better the local stellar mass
   function, it  over-predicts the clustering of  low-mass galaxies on
   small scales. The DLB07  model over-predicts the number of low-mass
   galaxies, but  it gives a reasonable  fit to the clustering  on small
   scales.  On larger scales ($10>r>1Mpc/h$), the predicted clustering in both models is  still   not perfec, but higher than the  data by around  30\%.

\item  We find that  the clustering  on small  scales is  dominated by
  satellite galaxies in the same  dark matter halo (so called one-halo
  term). The  one-halo term  is determined by  the density  profile of
  galaxies in the  host halo and the mass function  of host haloes. We
  find that  the  Guo11  model predicts a steeper  density profile of 
  galaxies. After using the same method to assign galaxy positions as that in the DLB07 model, the Guo11 model produces similar profiles as the DLB07 model, and the clustering on very small scales is suppressed. However, the discrepancy between the Guo11 and DLB07 models still exists at $r>0.1Mpc/h$. 

\item We compare the distribution of the host halo mass in the two models, and 
find that  there are
  slightly  more satellites residing  in massive  haloes in  the Guo11
  model. This  over-prediction of satellites in massive  haloes is the
  dominant contribution  to the discrepancy on galaxy clustering between the two models. Our  results do not support the  argument that the stronger clustering in  the Guo11 model is from the  formation bias of the host haloes (Wang13).


\item  We compare the  predictions on  the stellar mass functions in given halo mass from the DLB07 and Guo11 models,  and find  that both
  models over-predict the  number of satellites by the  same amount at
  $M_{\ast}=10^{10}M_{\odot}$ in massive haloes ($M_{vir}>2\times 10^{12}M_{\odot}$). The  DLB07  model  produces  more
  low-mass satellites than the Guo11 model. By simply  removing of 30\% of
  satellites in the two models,  they can both well fit the clustering
  data. Removal this  percent of satellites also  brings the Guo11
  model  into good  agreement with  the global  SMF. However,  for the
  DLB07 model we have to further remove about 60\% of central galaxies
  so as to fit  the global SMF.  By doing so, we  find that the total galaxy
  clustering is not suppressed on small scales, it is boosted instead.

\end{itemize}

We thus  conclude that the 'correct' prediction  of galaxy clusterings
from  the DLB07  model  is just  a  coincidence.  This  is because  it
predicts  too many  central galaxies  in low-mass  haloes,  which have
lower clustering.  The over-abundance  of centrals suppress the global
clustering of  all galaxies. However, we  note that our  simple way of
rescaling  the central galaxies  in the  DLB07 model  may not  be very
reasonable.   This is  because we  can not  simply throw  away central
galaxy arbitrarily,  unlike for  the satellites as  we can  argue that
current  consideration  of satellite  disruption  is  not included  or
inefficient. The right way of rescaling the DLB07 model is to move the
centrals into  host haloes with  higher mass, but lower  density.  For
example  for  central  galaxies  with  $M_{\ast}=10^{10}M_{\odot}$  we
should use  the positions of  central galaxies with higher  halo mass,
but with number  density about 60\% of the  current hosts. However, By
doing  so we  have to  also change  the host  of satellites  for given
stellar mass, this  is not possible as we do not  know how the stellar
mass of satellites evolve after their accretion in the DLB07 model.

Regarding the  Guo11 model, its main  problem is that  the fraction of
satellite galaxies  is higher  than the data  in massive  haloes. This
indicated  that the  tidal disruption  effect  in this  model is  less
efficient.  The Guo11  model can  thus  be improved  by introducing  a
slightly  stronger  tidal disruption  effect.  By  doing  so it  could
simultaneously  fit  the  global  stellar mass  function,  conditional
stellar  mass  function  in  different  host haloes,  and  the  galaxy
clustering on small scales.


We also  show in this  paper the semi-analytical  model of K12  with a
slight   modification   to  the   supernova   feedback  in   satellite
galaxies. In its previous version, K12  assume that the cold gas reheated by
supernova in  satellite galaxy is  reheated to the temperature  of its
host halo,  not the host  subhalo.  Here  we assume that  the cold  gas is
reheated to  the virial temperature  of its host subhalo.  Usually the
virial temperature of subhalo is lower than that of the massive host halo,
so  the   amount  of  heated  cold   gas  is  increased   in  the  new
description. We have found from the new model that,

\begin{itemize}
\item  The mass growth  for satellite  galaxies is  usually around a  
 factor of  2, much less than that in the old model, but  still slightly higher than the  data. We  find  that the CSMFs of  satellites now agree much  better with the
  SDSS data. We  note that the improvement in the  K12 model is purely
  due to the mass growth of satellites, not to the effect of satellite
  disruption. In most cases decreasing the mass of satellites
  has the same effect of satellite disruption. The reason is simple as
  there  is  good correlation  between  stellar  mass  and (sub)  halo
  accretion mass, and  the subhalo mass function is
  a power  law with negative index,  thus decreasing the  mass has the
  same effect of decreasing the number density.

\item The  clustering of  galaxies is now  well reproduced in  the new
  model, as there are fewer low-mass satellites in the new model.

\end{itemize}

Galaxy clustering  on small scales is  a hot topic  recently, being
seen as a common problem  of most semi-analytical models (e.g., Kim et
al. 2008;  Guo11; K12).  Many attempts have  been made to  solve this
problem,   including  adopting   a  low   $\sigma_{8}$   cosmology  or
introducing the warm dark matter model (Kang et al. 2013). However, it
is  found  that  $\sigma_{8}$  has  little  effect  on  the  predicted
clustering  (K12; Guo  et  al.   2013).  This  is  because  the decrease in dark matter clustering is almost entirely compensated by an increase in halo bias (Wang et al. 2008; Guo et al. 2013).

We have  clearly shown in this paper  that, for any kind  of model, if
one can simultaneously fit the global SMF and the CSMFs, combined with
a distribution  of galaxy density profile  like the NFW  one, we could
succeed  in  producing  the   clustering  on  all  scales.   Thus  for
semi-analytical models, it can be achieved by constraining the physics
governing satellite evolution, such as mass stripping and disruption.

\section{Acknowledgements}

Xi Kang thanks  Simon White for helpful discussions,  and Ramin Skibba
for useful  comments and  careful reading of  the manuscript.  We also
thank  the referee  for  useful comments.   The Millennium  Simulation
databases used in this paper  and the web application providing online
access  to them  were constructed  as part  of the  activities  of the
German  Astrophysical   Virtual  Observatory  (GAVO).   This  work  is
supported   by  the   National   basic  research   program  of   China
(2013CB834900), and  the NSFC (No.  11073055, 11333008). Xi  Kang also
acknowledges  the  support from  the  Bairen  program  of the  Chinese
Academy   of   Sciences   and    the   Partner   group   between   the
Max-Planck-Institute   for   Astronomy   and   the   Purple   Mountain
Observatory.  The  simulation is run  on the Supercomputing  center of
CAS.

\bibliographystyle{mn2e}
\bibliography{paper.bbl}


\label{lastpage}
\end{document}